\documentclass[structabstract]{aa}  
\usepackage{graphicx}
\usepackage{rotating}
\usepackage{pdflscape}
\usepackage{txfonts}

\usepackage{natbib}
\usepackage{enumerate}
\def\feii{log n(Fe~{\sc II)}}
\def\fei{log n(Fe~{\sc I)}}
\def\logg{$\log$~g}

\bibpunct{(}{)}{;}{a}{}{,} 

\begin{document}

   \title{{\sc{FAMA}}: An automatic code for stellar parameter and abundance determination\thanks{The current version of FAMA is available at the CDS via anonymous ftp 
   to cdsarc.u-strasbg.fr (130.79.128.5) or via  http://cdsweb.u-strasbg.fr/cgi-bin/qcat$?$J/A+A/}}


\author{Laura Magrini\inst{1}, Sofia Randich\inst{1}, Eileen Friel\inst{2}, Lorenzo Spina\inst{1,8}, Heather Jacobson\inst{3}, Tristan Cantat-Gaudin\inst{4,5}, Paolo Donati\inst{6,7}, Roberto Baglioni\inst{8}, Enrico Maiorca\inst{1}, Angela Bragaglia\inst{6}, Rosanna Sordo\inst{4}, Antonella Vallenari\inst{4}}

\institute{
INAF--Osservatorio Astrofisico di Arcetri, Largo E. Fermi, 5, I-50125 Firenze, Italy
\email{laura@arcetri.astro.it}
\and
Department of Astronomy, Indiana University, Bloomington,  USA
\and
MIT Kavli Institute, Boston, USA
\and
Osservatorio Astronomico di Padova, Vicolo dell'Osservatorio, 5, 35122, Padova, Italy
\and
Dipartimento di Fisica e Astronomia, Vicolo dell'Osservatorio, 3, 35122, Padova, Italy
\and
INAF-Osservatorio Astronomico di Bologna, Via Ranzani, 1, 40127, Bologna, Italy
\and
Dipartimento di Fisica e Astronomia,  Via Ranzani, 1, 40127, Bologna, Italy
\and
Dipartimento di Fisica, sezione Astronomia, Largo E. Fermi, 2, 50125, Firenze, Italy
}

   \date{Received ; accepted }

 
  \abstract
{The large amount of spectra obtained during the epoch of extensive spectroscopic surveys of Galactic stars
needs the development of automatic procedures to derive their atmospheric parameters and individual 
element abundances.  }
{Starting from the widely-used code  MOOG by C. Sneden, we have developed
a new procedure to determine atmospheric parameters and abundances in a fully automatic 
way. The code FAMA (Fast Automatic MOOG Analysis) is presented describing its approach 
to deriving atmospheric stellar parameters and element abundances. The code, freely distributed, is  
written in Perl and  can be used on different platforms.  }
{The aim of FAMA  is to render  the computation of 
the atmospheric parameters and abundances of a large number of stars using measurements of  equivalent widths as automatic and as independent of any subjective approach as possible.  It is based on the simultaneous search for three equilibria: 
excitation equilibrium, ionization balance, and the relationship between \fei\ and the reduced equivalent widths. 
FAMA also evaluates the statistical errors on individual element abundances and errors due to the uncertainties in the stellar parameters.  
The convergence criteria are not fixed 'a priori' but are based on the quality of the spectra.  
   }
{In this paper we present tests performed on the Solar spectrum EWs which  tests the dependency on the initial parameters, and  the analysis  of a sample of stars observed in Galactic open and globular clusters.
}
{}
\authorrunning{Magrini, L. et al.}
\titlerunning{\sc FAMA}

   \maketitle

\section{Introduction}

The most recent  years have been  characterized by a large number of ambitious spectroscopic surveys, 
such as,  
the {\sc Gaia-ESO Survey (GES)}  with the  Very Large Telescope (VLT) at the European Southern Observatory \citep{gilmore12, rg12}, the {\sc APOGEE} Survey at the Apache
Point Observatory \citep{allende08}, the {\sc RAVE}
Survey at the Anglo-Australian Observatory \citep{zwitter08}, and the  {\sc BRAVA}  
\citep{kunder12} and {\sc ARGOS} \citep{freeman13} surveys dedicated to the Galactic bulge. 
Moreover, wide-field multifiber facilities  such as,  {\sc HERMES} \citep{barden10} at the Anglo-
Australian Observatory, {\sc 4MOST} \citep{dejong11} at the New
Technology Telescope, and {\sc MOONS} \citep{cirasuolo11}
at the VLT are planned for the near future.

These surveys yield spectra for a very large number of stars which need automatic or semi-automatic procedures to analyze their spectra. 
To satisfy the requests for an automatic and rapid analysis, several new codes have been designed during the past few years. 
Some of them are based on the $\chi^2$ minimization of the differences between the observed spectrum 
and a grid of  synthetic spectra, such as the codes  {\sc ABBO} \citep{bonifacio03} and  
SME \citep{valenti96}, or the  projection of the observed spectra on a grid of synthetic spectra as in MATISSE \citep{recioblanco06}. Other methods are based on the analysis of the equivalent widths of the metal lines, as in  the 
codes developed by Takeda and collaborators \citep{takeda02} and  in {\sc GALA} \citep{mucciarelli13}. 

The two methodologies, the global   $\chi^2$ minimization  and the traditional analysis with EWs, differ particularly 
in their use of the knowledge  of the physics, the Boltzmann and Saha statistical  physics,  governing the stellar atmospheres. 
While  classical methods make an intensive use of statistical  physics,  taking advantage of the physical-state      
indicators that  are available in high-resolution stellar spectra, the  $\chi^2$ minimization methods lose part of the information 
trying to optimize the atmospheric parameters by seeking the best match between the observed spectrum and a synthetic spectrum.
The possible risk of this latter approach is, for instance, described by 
 \citet{torres12} who analyzed a sample of high-resolution spectra with both the classical EW method and the $\chi^2$ minimization 
techniques. Comparing  the results  obtained with different methods, they found that correlations between the stellar parameters exist,  
and that they are much stronger for parameters derived with   $\chi^2$ minimization 
due to one spectroscopic quantity that can play against another to some extent  which leads to nearly the same cross-correlation value or  $\chi^2$ value. 
For example, stronger lines produced by adopting a higher metallicity for the synthetic spectrum can be compensated by a suitable increase in the effective temperature to first order. 
A similar degeneracy exists between surface gravity and temperature. 
\citet{torres12} found that the effect of these correlations is evidently much weaker in the case of classical EW analysis, in which their work was performed with {\sc Moog}. 
For these reasons we believe that traditional analysis has still a strong potential in the framework 
of high-resolution spectra, and, with the help of automatic procedures, it will be competitive as one of the more appealing    $\chi^2$ minimization  techniques.

In this context, we have decided to automate the well-known,  widely-used  {\sc Moog} code\footnote{{\sc Moog} is available at http://www.as.utexts.edu/$\sim$chris/moog} \citep{sneden73, sneden12} 
to allow the computation of a fully self-operating spectral analysis. 
{\sc Moog} was originally designed at the beginning of the 1970s, and it has been periodically updated 
by Sneden and his collaborators.
The typical use of {\sc Moog} is to assist 
in the determination of the chemical composition of a star, following the  basic equations of one dimensional local thermodynamic equilibrium (1D LTE)  stellar line analysis. 
{\sc Moog}   performs a variety of  line analysis and spectrum synthesis tasks.  

The classical version of {\sc Moog} makes use of the {\sc Abfind} driver with a model atmosphere to derive the  stellar parameters and  the abundances. {\sc Moog}  needs a continuous human intervention to adjust the stellar parameters satisfying  
three requirements: the ionization balance, the excitation balance, and the minimization of the  trend between \fei\ and the observed (or predicted)  reduced equivalent widths  [log(EW/$\lambda$)].
Each of these three well-known steps allows one to fix one of the stellar parameters: 
The effective temperature
(T$_ {\rm eff}$) is obtained by eliminating  trends between \fei\,
and the excitation potential (EP). The surface gravity is optimized by
assuming the ionization equilibrium condition, i.e. \feii=\fei. The microturbulence is set by minimizing the slope of the
relationship between \fei\, and the logarithm of the  reduced EWs.
However the three steps are interconnected, for example  a variation in effective temperature produces a variation 
in  \feii\, and \fei and thus could require an adjustment in¶ the surface gravity. 
The changes on gravity can sometimes be of second order. As long as changes to T$_ {\rm eff}$ and $\xi$ are not radical from iteration to iteration, 
we can thus neglect the variation in $\log$ g by adjusting it every few iterations. 
The determination of the atmospheric parameters is usually done with an iterative process, which could take a relatively long time to complete  the analysis of a single star. 
This process is acceptable, if the number of stars is limited, but it is completely infeasible for a large survey.  
 
To make the analysis of a large number of stars possible, we have designed 
an automation of  {\sc Moog}, also called {\sc FAMA} (Fast Automatic Moog Analysis),  that does not need any human intervention during the phase of determination of abundances and
atmospheric parameters. It is able to minimize the slopes of the correlations in a coherent way that considers  the quality of the spectra and of the EW measurement. 
{\sc FAMA} is  developed at the Arcetri Observatory and it can be freely obtained  with the installation manual sending an email to {\tt laura@arcetri.astro.it}, in the 
web page {\tt http//www.arcetri.astro.it/$\sim$laura}, or at CDS. 
It is a  Perl code\footnote{For {\sc FAMA} we used Perl, but the same steps can be obviously done in other languages, such as  Fortran or Idl}, 
 which performs a complete 1D LTE (local thermodynamic equilibrium) spectral analysis of high-resolution spectra. Composed of several modules and a parameter file that allow the user to derive the stellar parameters and abundances, the Perl code
starts from two files with EWs (one with iron EWs and the other with EWs of other elements) and a file with first-guess atmospheric parameters. 

In the present paper we show the working principles of {\sc FAMA} 
with some tests on the Sun and on a sample of stars in open clusters (OCs) and in globular clusters (GCs). 
The paper is structured as follows: In Sec.2 we present the philosophy of  the code together with the definition of the quality parameter {\sc QP}. 
In Sec.3, we will describe how the errors are computed, whereas in Sec.4 we present the test 
data analysis. Finally, our summary and conclusions are given in Sec.5.


\section{The working principles of the code} 

The ingredients with which {\sc FAMA} is fed  are: i) {\em EW files}:  two files, one containing EWs of iron in the two ionization stages and 
the second one  containing EWs belonging to the complete list of elements, and ii) {\em parameter file}: a file with the first-guess atmospheric parameters which includes the effective temperature,  T$_{\rm eff}$, 
the surface gravity, \logg\,, the microturbulent velocity, $\xi$, and the abundance of iron with respect to the solar value, [Fe/H]\footnote{[Fe/H]=\fei - \fei$_{\odot}$}. 

The  philosophy of the code is based on an iterative search for the three equilibria (excitation, ionization, and the trend between \fei\ and log(EW/$\lambda$)) 
with a series of recursive steps starting from a set of initial atmospheric parameters, and arriving at a final set 
of atmospheric parameters which fulfills the three equilibrium conditions.
The order followed in the search for the three equilibria  is also important since T$_{\rm eff}$ is the controlling parameter for the ultimate solution. Thus it is necessary 
first to regulate it, then to move to the second most important parameter, the surface gravity, which adjusts the ionization equilibrium, and 
finally to fix  the microturbulence.   

\begin{itemize}
\item[1.] {\em Excitation equilibrium}: \\
According to the classical equation of Saha-Boltzmann, as summarized  by 
\citet{takeda02} and by  \citet{gray05}, 
the abundance of  Fe~{\sc I} increases with an increase in T$_{\rm eff}$, while it is less
sensitive to a variation in the surface gravity.  
The relation is given by the following formula:
\begin{equation}
{\rm log~n} ({\rm FeI}) \propto e^{-EP_{i}/kT_{\rm eff}},
\end{equation}
where \fei\ is the logarithmic abundance of iron and EP$_{i}$ is the excitation potential of each line corresponding to the level $i$. This relation is valid in the approximation that most of the iron atoms  are neutral.  
This means that the variation in temperature causes changes in iron abundance of lines with different  EP$_{i}$, and we can constrain 
T$_ {\rm eff}$  by requiring that  the abundances derived from Fe~{\sc I} lines, for which
the values of the excitation potential spread over a sufficiently
wide range, be independent of EP. 
\item[2.] {\em Ionization equilibrium}:\\
The surface gravity, \logg, is related to the  Fe~{\sc II} abundance by the following relation
\begin{equation}
{\rm log~n} ({\rm FeII})\propto \log g^{n/3}, 
\end{equation}
derived in the approximation of the classical equation of Saha-Boltzmann and with n$=1$  when iron is mostly ionized and n$=2$ for an iron population dominated by neutral atoms. 
Thus, the same abundance should be derived 
for neutral and  ionized lines in the condition of ionization equilibrium for a given element. The equality of the abundances \fei\ and \feii\ gives us a direct measurement 
of the surface gravity.

\item[3.] {\em Microturbulence equilibrium}: \\
The minimization of the trend  between \fei\  and the reduced equivalent widths, log(EWs/$\lambda$),  are necessary to set the microturbulent velocity, $\xi$. 
The microturbulent velocity is a quantity incorporated in the Doppler width of the profile of a line, which was introduced 
to increase the strength of the lines near or on the flat part of the curve of growth to reconcile them with the observations.  
The classical way to set $\xi$ is to nullify any
trend between the abundance  \fei\  and the reduced equivalent
width. The justification for this requirement  is that  $\xi$  preferentially  affects the moderately/strong
lines, while the weak ones are less affected by $\xi$.  
As noted by \citet{magain84},   this classical method might lead to a systematic overestimate of $\xi$, when the observed EWs are 
affected by random errors. This overestimate is due to the correlation between errors in EWs and in abundances 
of each line. This might be avoided by using theoretical EWs instead of observed ones.

\end{itemize}

\begin{figure*}
\centering
\includegraphics[width=1\textwidth]{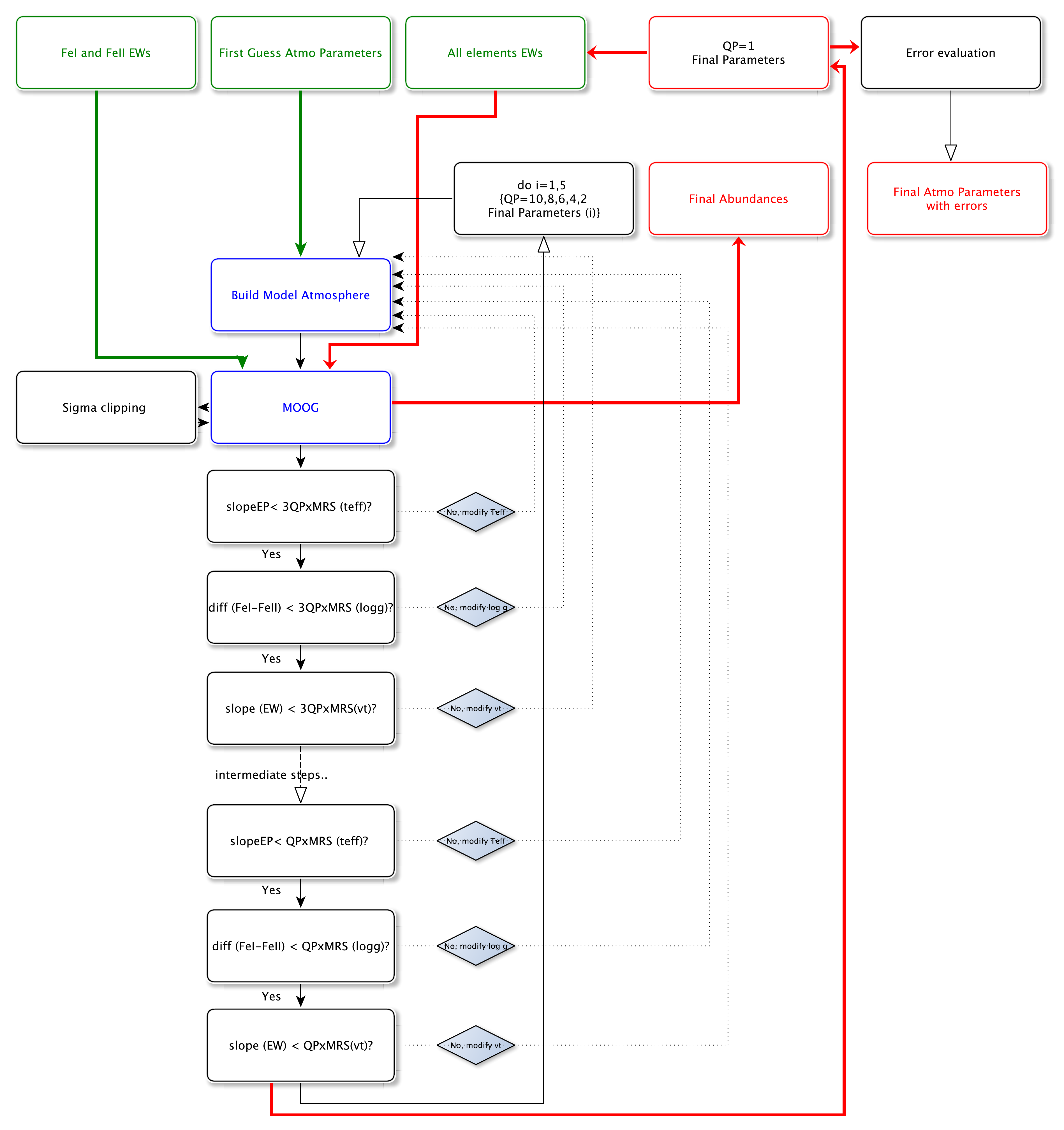}
\caption{Block diagram of {\sc FAMA}: the input files in green,  the main codes in blue, the iterative steps in black, and finally, 
the final outputs in red. }
\label{block}
\end{figure*}

Thus, the aim of {\sc FAMA} is to reach the three equilibria with a series of iterations. 
A block diagram of {\sc FAMA} is shown in Fig.~\ref{block}. 
The iteration strategy is the following: First the T$_{\rm eff}$ is adjusted by an amount that depends on how far the initial  T$_ {\rm eff}$ is  from the excitation equilibrium. 
Subsequently the surface gravity is modified by an initial amount that depends on the difference between \fei\ and \feii.
Finally, $\xi$  is varied on the basis of the slope of reduced EW versus \fei.    
This is repeated in three cycles (see the first and last cycle in Fig.~\ref{block}). In each cycle, the minimization requirements on the slopes and neutral/ionized iron abundances are varied and becomes stricter with each cycle. In particular,   
the minimization requirements for the first cycle are three times larger than those of the last cycle, and those of the second cycle are two times larger.  
The value of the smallest minimization requirement is calculated using the information on the quality of the EW measurements.  
Since the EW measurements are affected by errors, it is not reasonable to minimize the slopes to infinitely low values, yielding zero slopes. 
This would have no physical justification and would lead us to find local minima in the three-dimensional space of T$_{\rm eff}$, \logg, and $\xi$.  
Thus the minimum reachable slopes (MRS) in {\sc FAMA}  are strictly linked to the quality of the spectra, as expressed by the dispersion 
of \fei\ ($\sigma_{\rm FeI}$) around the average value $<$ \fei$>$. This is correct in the approximation that the main contribution to the dispersion is due to 
the error in the EW measurement rather than to an inaccuracy in atomic parameters, as e.g.,  the strengths ($\log$~gf).
The MRS are defined as the following: 
\begin{itemize}
\item[i)] MRS$_{\rm T_ {\rm eff}}$=$\frac{(\sigma_{\rm FeI}/\sqrt(N_{\rm lines})}{rEP}$, where rEP is the range spanned in excitation potential (EP) by the measured lines and ${(\sigma_{\rm FeI}/\sqrt(N_{\rm lines})}$ is the standard deviation with N$_{\rm lines}$ the number of  lines employed;  
\item[ii)] MRS$_{\xi}$=$\frac{(\sigma_{\rm FeI}/\sqrt(N_{\rm lines})}{rEW}$, where rEW is the range spanned in reduced EW;
\item[iii)] MRS$_{\rm log g}$=$\sqrt{\sigma_{\rm FeI}^{2} + \sigma_{\rm FeII}^{2}}$, where $\sigma_{\rm FeI}^{2}$ and $\sigma_{\rm FeII}^{2}$ 
are the dispersions around the mean of \fei\ and \feii, respectively.  
\end{itemize}
This means that we are allowed to minimize the excitation equilibrium up to a  MRS$_{\rm T_ {\rm eff}}$$\sim$0.001 for a spectrum with a very high signal-to-noise ratio (SNR) having a $\sigma_{\rm FeI}$=0.06~dex derived from about one hundred Fe~{\sc I} lines, 
and a typical rEP=5, while  we cannot reach MRS$_{\rm T_ {\rm eff}}$$\leq$0.003 for a spectrum of average quality with a $\sigma_{\rm FeI}$=0.12~dex, derived from $\sim$60 lines. 
The same is true for the ionization equilibrium and for the microturbulent velocity trend.

During the first iteration, the file containing the EWs of  Fe~{\sc I} and  Fe~{\sc II} are treated with the 
$\sigma$-clipping, a procedure that 
allows us to remove EWs producing abundances 
more discrepant than $n\sigma$ from the average abundance.
The $\sigma$-clipping thresholds may be easily changed on the basis of the needs of the user and of the quality and spectral coverage of the spectrum via a file containing the main parameters that can be set by the user. 

\subsection{The {\sc QP} parameter} 

The first-guess atmospheric parameters should be derived from photometric information or from other sources, if possible, and should preferably allow one to start close to the final values. 
That is a dwarf star should 
be treated with the initial parameters suitable for a dwarf star, and a giant star should have appropriate first guess parameters.    
To ensure the final solution independence of the initial parameters, {\sc FAMA} is however designed to repeat  the complete convergence path up to six times, 
starting each time from the previous set of stellar parameters which ensured the convergence. 
At each step, the requirements of minimization of the three slopes become stricter and they are parametrized by the so-called {\sc QP} quality parameter.
This parameter {\sc QP}  assumes six values, 10, 8, 6, 4, 2, and 1, which means that the requirement for the first step is that, e.g.,  a MRS$_{\rm T_ {\rm eff}}$=10$\times$$\frac{(\sigma_{\rm FeI}/\sqrt(N)}{rEP}$.  In
the specific example described above, the  high SNR spectrum would indicate  MRS$_{\rm T_ {\rm eff}}$$\sim$0.01.  In the later steps, MRS$_{\rm T_ {\rm eff}}$ will move from 0.01 to 0.001, 
passing from 0.008 to 0.006, 0.004, and 0.002.  Note that each step starts with initial parameters that are the convergence point of the previous step and with the {\em original} EW list. 
  The sequence of several steps is necessary, because it allows us  to move toward the final set of stellar parameters while maintaining the initial line list. 
If we were to reach a minimization of the slope with {\sc QP}=1 with a single  step, by starting from  stellar parameters that were far from the true ones, we would risk both losing several lines with the first 
$\sigma$-clipping and finding  a convergence point that could be far from the true value.  
The values assumed by the {\sc QP} parameter can be set by the user on the basis of the quality of spectra.

From the block diagram of {\sc FAMA} shown in Fig.~\ref{block} we can see the structure of the code. 
At the top of the diagram, in green, are the three input files. 
The flow of the analysis is the following: The file containing the first guess stellar atmospheric parameters is given to an interpolator code of model  atmosphere, in our specific case the KURUCZ \citep{kurucz79,ck04} or MARCS models \citep{gustafsson08}, and 
a model atmosphere with the input parameters is produced. 
The model atmosphere and the files containing the EWs of iron are given to {\sc Moog}.  A $\sigma$-clipping is performed on  Fe~{\sc I} and  Fe~{\sc II} lines based on an initial run of {\sc Moog}, and 
{\sc Moog} is run again with the cleaned list of EW. 
Then, the first step begins  with the {\sc QP} parameter set to 10.  Each step is composed of a series of iterations (shown with dotted arrows), which allow {\sc FAMA} to reach the required  minimization criteria. 
The iterations continue  until  the final convergence 
criteria are reached. The last step 
(indicated with red lines and arrows) corresponds to {\sc QP}=1. At the end of this step when the final stellar parameters are obtained, the EWs of all elements are given to {\sc Moog}, a $\sigma$-clipping 
is then performed on them, and a final evaluation of the error is done. At this point, the final results are written in two files, one containing the final stellar parameters with errors, 
and the other with the individual element abundances and their errors (in red in the diagram).  

\subsection {Quality plots} 
\begin{figure*}
\centering
\includegraphics[width=1.0\textwidth]{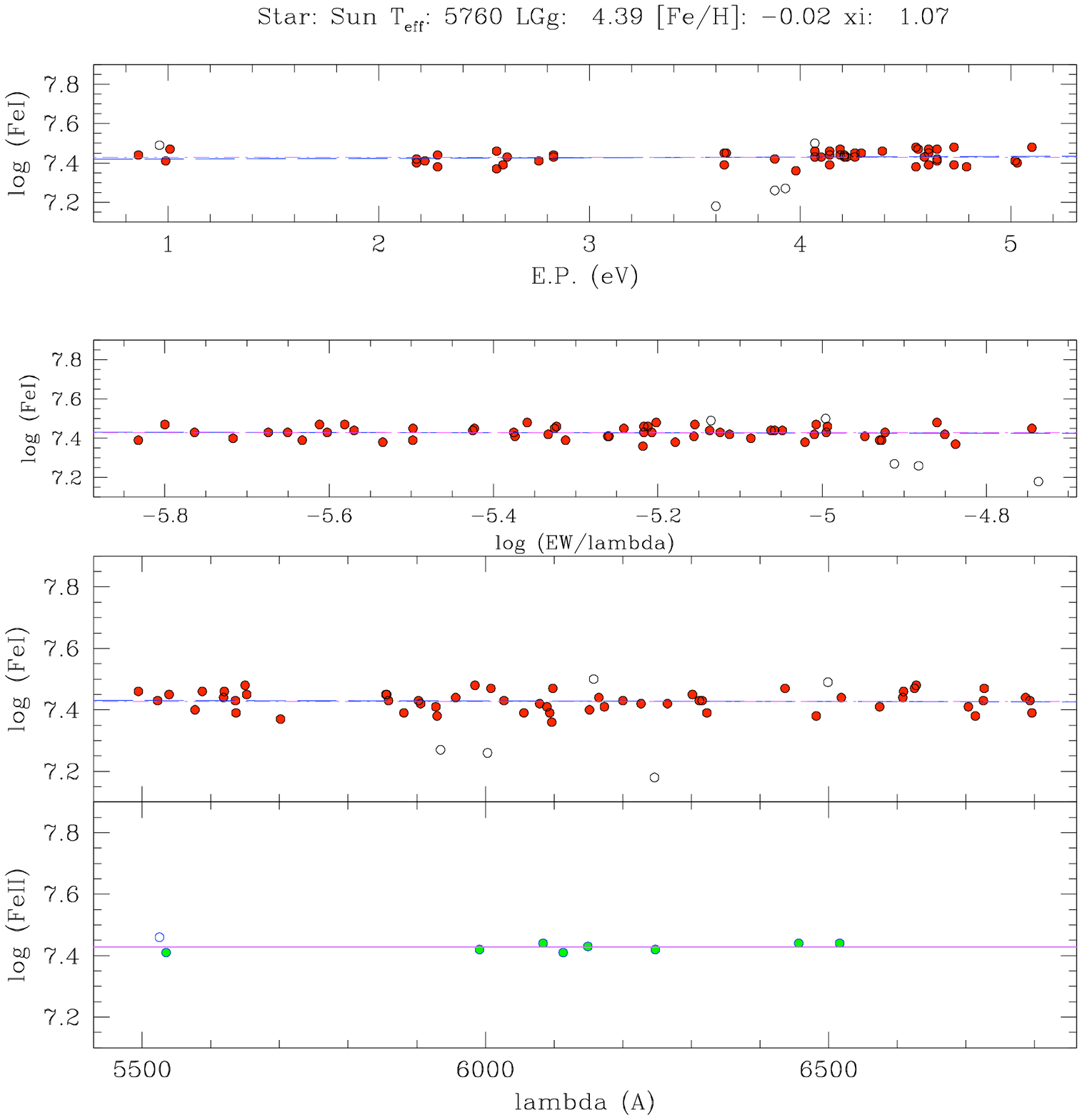}
\caption{Example of the final control plot of {\sc FAMA}.  In the first three panels,  the filled (red) circles are the abundances from the Fe~{\sc I} EWs accepted after the $\sigma$-clipping, while the empty circles are those rejected. 
In the first two panels the excitation and microturbulence equilibria are shown. The blue long-dashed line is the zero-slope curve, and the dashed magenta line is the slope of the final convergence point. 
In the third panel the dependence on iron abundances on $\lambda$ is shown, and in the fourth panel,  the ionization equilibrium is presented with the green circles being  
the Fe~{\sc II} abundances used for gravity determination. 
The cyan horizontal line in the last panel is the average Fe~{\sc II} abundance, while the magenta line is the average Fe~{\sc I}.  
}
\label{fig_control}
\end{figure*}

At the end of the run of {\sc FAMA} a control plot is produced with the aim of helping to  visually check  the quality of the result\footnote{The quality plots are produced making use of a {\sc supermongo} macro {\tt http://www.astro.princeton.edu/$\sim$rhl/sm/}. }. 
In Fig.~\ref{fig_control}, we show an example of the control plot with four panels. In the first three panels, the filled (red) circles are the abundances from the  Fe~{\sc I} EWs accepted after the $\sigma$-clipping, while the empty circles are those rejected. 
In the first two panels, the excitation and microturbulence equilibria are shown. The blue long dashed line is the zero-slope line, and the dashed magenta line is the slope of the final convergence point. In the case of good convergence these two lines are coincident. 
In the third panel, the dependence on iron abundances on $\lambda$ is shown giving us a further test on the quality of the data and of the EW measurements. In spectra of good quality, we do not expect any dependence of EWs on the wavelength. 
Finally,  the ionization equilibrium is presented in the fourth panel  with the green circles, where the  Fe~{\sc II}  abundances are used for gravity determination, and  with empty circles  showing rejected lines.
The cyan horizontal line in the last panel is the average  Fe~{\sc II} abundance, while the magenta line is the average  Fe~{\sc I}.

\section{The evaluation of errors}
\begin{figure}
\centering
\includegraphics[width=0.55\textwidth]{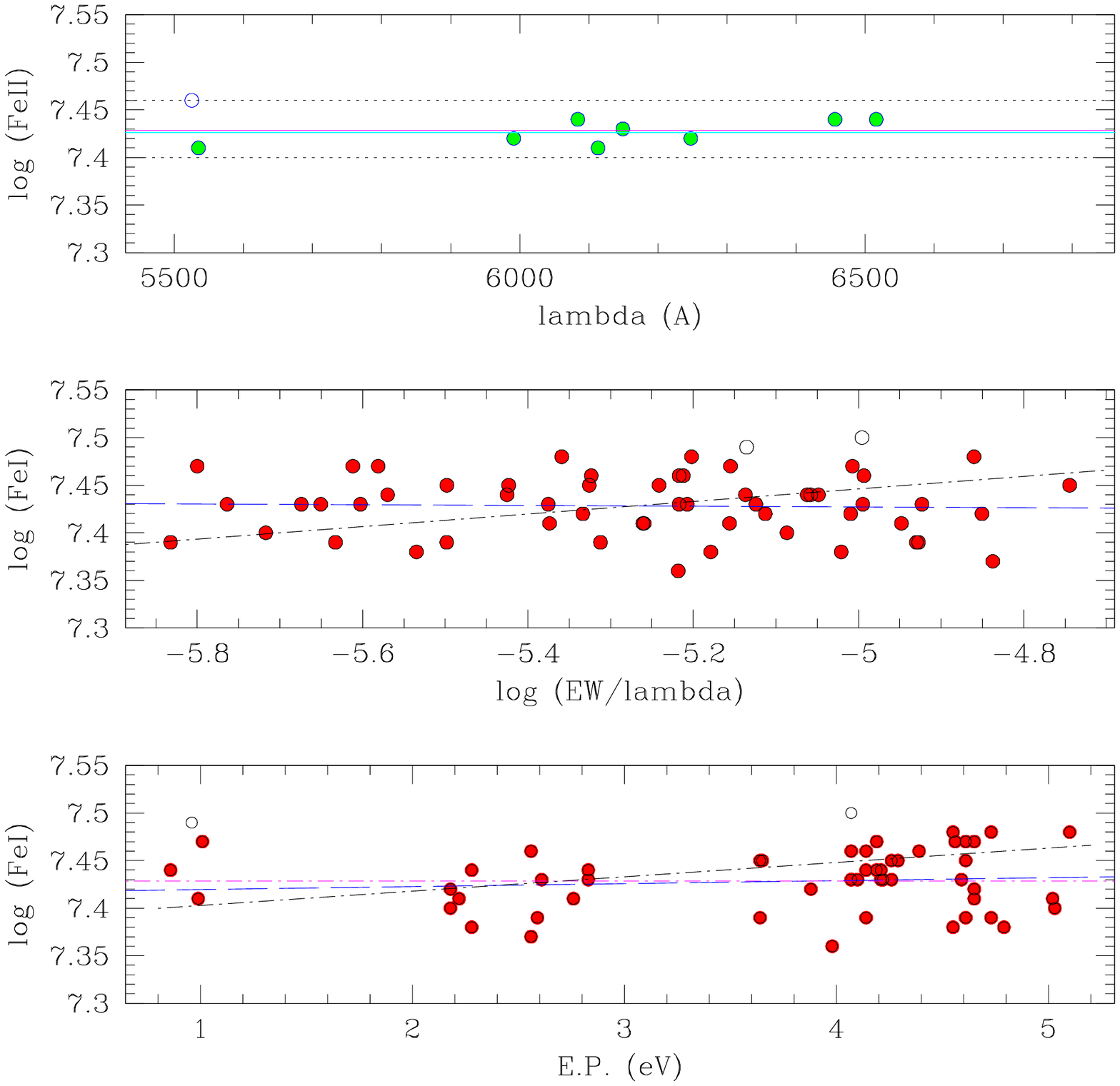}
\caption{Evaluation of errors on the stellar parameters.  The green circles are the lines accepted after the $\sigma$-clipping.
The blue long-dashed line is the zero-slope curves, and the dashed magenta line is the slope of the final convergence point. 
The two dotted lines in the upper panel indicate $\pm$$\sqrt{\sigma_{\rm FeI}^{2} + \sigma_{\rm FeII}^{2}}$ values. The dashed-dotted lines 
in the second and last panels show  the lines with a slope of $\frac{\sigma_{\rm FeI}}{rEP}$ and of $\frac{\sigma_{\rm FeI}}{rEW}$, respectively. }
\label{fig_sun_err}
\end{figure}

{\sc FAMA} is designed to evaluate the errors on the final stellar parameters. 
This is done with a further last step after the final convergence, in which the slopes of the excitation  equilibrium, the slope of the trend between  \fei\ and reduced EWs, 
and the difference between  Fe~{\sc I} and  Fe~{\sc II} abundances  are not completely minimized but kept to the values given by the dispersion of the abundances. 
To do this, {\sc FAMA} derives  the stellar parameters which correspond  to following: \\
1) For the T$_{\rm eff}$:  a slope equal to the ratio between the dispersion around the mean value of iron abundance and the range of EP  $\pm$$\frac{\sigma_{\rm FeI}}{rEP}$; \\
2) For $\xi$: a slope corresponding to $\pm$$\frac{\sigma_{\rm FeI}}{rEW}$, where rEW is the range spanned in reduced EW; \\
3) For gravity:  a difference between \fei\ and \feii\ equal to $\pm$$\sqrt{(\sigma_{\rm FeI}^{2} + \sigma_{\rm FeII}^{2})}$, where $\sigma_{\rm FeI}^{2}$ and $\sigma_{\rm FeII}^{2}$ 
are the dispersions around the mean of \fei\ and \feii, respectively.  
This allows us to find  the maximum  and minimum values for T$_{\rm eff}$, \logg, and $\xi$,  which are acceptable within errors due to the dispersion of the abundances. 
This choice is shown in Fig.~\ref{fig_sun_err}, where the slopes corresponding to the dispersion of the Fe~{\sc I} and Fe~{\sc II} abundances are shown. 

Concerning  the error in the final individual element abundances, there are two types of errors  which are considered  by {\sc FAMA}: {\em i)} the statistical uncertainties due to the random errors 
in the EW measurements and to uncertainties on the atomic parameters; and {\em ii)} the errors  on the abundances generated by  the uncertainties in the determination of the atmospheric parameters. 
The first source of errors is simply evaluated during the analysis with {\sc Moog} by the standard deviation around the average value of the abundance of each elements. 
The second source of error is instead evaluated recomputing the element abundances with the 
stellar parameters corresponding to the minimum and maximum slope and Fe~{\sc I} and Fe~{\sc II} difference. 
Their difference with respect to the abundances computed with the best parameters gives us  an estimate of the error due to the uncertainties in stellar parameters.

\section{Code validation}

The aim of the present paper is to introduce the {\sc FAMA} code testing it on a set of literature EW measurements. 
We have used the EWs of the Sun and of a set of stars in OCs and GCs present in published papers to perform a set of tests 
described in the following sections.   

\subsection{Checking the independency of the starting point with the Sun}

\begin{figure}
\centering
\includegraphics[width=0.5\textwidth]{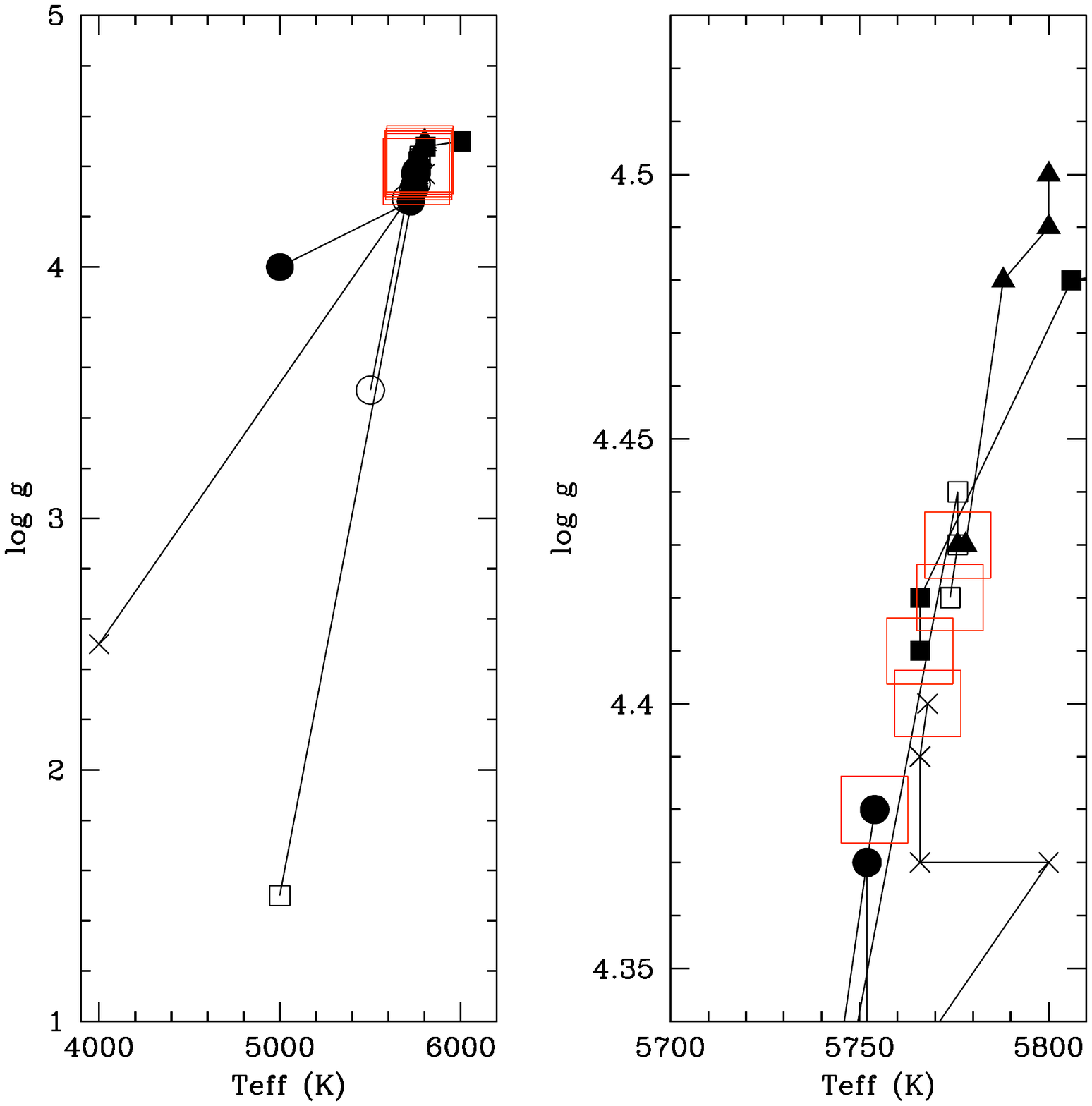}
\caption{Solar parameters derived from a sample of different first guess stellar parameters. The different symbols indicate the different initial parameters shown in Table~\ref{tab_sun}. The (red) empty large squares are the final parameters shown in the last four columns of Table~\ref{tab_sun}.   }
\label{fig_sun}
\end{figure}

First, we have considered the EWs of the Sun measured by Magrini et al. (2010. hereafter M10) from a high-resolution spectrum that is obtained with the spectrograph UVES at VLT. 
We have performed the analysis with {\sc FAMA} starting from six different sets of atmospheric parameters. 
The aim of this first test is to check the possibility of retrieving   with 
initial parameters that stay very far from the known parameters. 
In Table~\ref{tab_sun}, we show the results of {\sc FAMA}, with the literature accepted solar values: All results are in excellent agreement with the Solar parameters. 
In Fig.~\ref{fig_sun} the paths followed by {\sc FAMA}  are shown. The left panel shows the full range of T$_ {\rm eff}$ and \logg\, while the right panel is a zoom 
around the region of the final solutions.   
Different symbols indicate the path obtained from each of the six starting values of Table~\ref{tab_sun}. 
The red squares mark the final values. Note that {\sc FAMA} is already approaching the final values  after the first step. 
The following steps allow us to  fine-tune the parameters and to approach as close as possible the final values. 
The uncertainties shown in Table~\ref{tab_sun} are obtained with the procedure described in Sec.~3, and the error in 
the iron abundance considers both sources of errors from EWs and stellar parameters. 
We note that all solutions are consistent within error\footnote{No code is able to produce atmospheric parameters  more accurate than those permitted by the data uncertainties.} of the accepted solar parameters given in the first row of the Table~\ref{tab_sun}.

\begin{table*}
\begin{center}
\caption{Solar analysis. }
\tiny
\begin{tabular}{llllllll}
\hline\hline
T$_{\rm eff-in}$ & log g$_{\rm in}$ & [Fe/H]$_{\rm in}$ & $\xi_{\rm in}$ &  T$_{\rm eff-out}$ & log g$_{\rm out}$ & [Fe/H]$_{\rm out}$ & $\xi_{\rm out}$\\
\hline\hline
 \multicolumn{4}{c}{Literature  Solar parameters} & 5777  & 4.44   & 0.0	& 1.1 \\	
\hline \hline
5500 & 3.51 & 0 & 0.9	&   5754$\pm$81 & 4.38$\pm$0.19 & -0.02$\pm$0.07 & 1.07$\pm$0.05 \\
5000 & 4.01 & 0 & 1.0	&   5754$\pm$81 & 4.38$\pm$0.19 & -0.02$\pm$0.07 & 1.05$\pm$0.12 \\
6000 & 4.51 & 0 & 0.8	&   5766$\pm$100 & 4.41$\pm$0.20 & -0.01$\pm$0.09  & 1.05$\pm$0.07 \\
4000 & 2.51 & 0 & 1.2	&   5768$\pm$94 & 4.40$\pm$0.20 & -0.02$\pm$0.08 & 1.07$\pm$0.06 \\
5000 & 1.51 & 0 & 1.5	&   5774$\pm$95 & 4.42$\pm$0.20 & -0.01$\pm$0.08 & 1.05$\pm$0.06\\
5800 & 4.51 & 0 & 1.1	&   5776$\pm$118 & 4.43$\pm$0.20 & -0.01$\pm$0.08  & 1.05$\pm$0.06 \\
     \hline
\hline
\end{tabular}
\label{tab_sun}\\
\end{center}
\end{table*}

\subsection{Stars in open clusters}
We have selected a sample of stars belonging to OCs, observed by several authors during the past few years and analyzed with {\sc Moog}. The sample includes  
nine evolved stars in the old, distant open clusters Be18, Be21, Be22, Be32 , and PWM4 (Yong et al. 2012,  hereafter Y12), 16 evolved stars in the old open clusters 
Be 31, Be 32, Be 39, M 67, NGC 188, and NGC 1193  (Friel et al. 2010,  hereafter F10), and 10 stars \citep{magrini10} in three inner-galaxy  clusters, NGC 6192, NGC 6404, NGC 6583. 
We have preferred to use literature measurements to separate the effects due to EWs and the choice of atomic parameters from the effects due 
to minimization criteria. Thus,  we use the  Fe~{\sc I} and Fe~{\sc II} EWs available in the original papers and the atomic data adopted by each author for each star.
The aim of this paper is indeed to compare the {\sc Fama} results to those of the literature which are obtained in the same conditions (same atomic parameters and EWs). 
A future work will be devoted to a re-analysis using a common line list and a common method of measuring EWs and deriving parameters.

This test focuses on  stars analyzed with the classical procedure. That is the author using the classical procedure start from 
the photometric values of T$_{\rm eff}$ and \logg\ and from an assumed  $\xi$  1.5 km s$^{-1}$ for giant stars, and manually adjust effective 
temperatures to remove any trends of Fe~{\sc I} abundance with EP.  Then, they alter  \logg\ values to achieve ionization equilibrium, vary $\xi$ to minimize trends of Fe~{\sc I} abundance with line strength and iterate as necessary on the parameters until the trends are removed. 
Starting from the same photometric values we have performed the analysis on the EW values published in the paper of Y12, F10, and M10, with also their EP and log~{\em gf} values. 
In our analysis we have considered  the line broadening  computed with tabulated damping constants \citep{barklem00, barklem05} whenever possible, and we have  used the Uns\"old approximation for other lines. 
In the works of M10, F10 and Y12, the Uns\"old approximation has been adopted. 

\begin{figure*}
\centering
\includegraphics[width=0.9\textwidth]{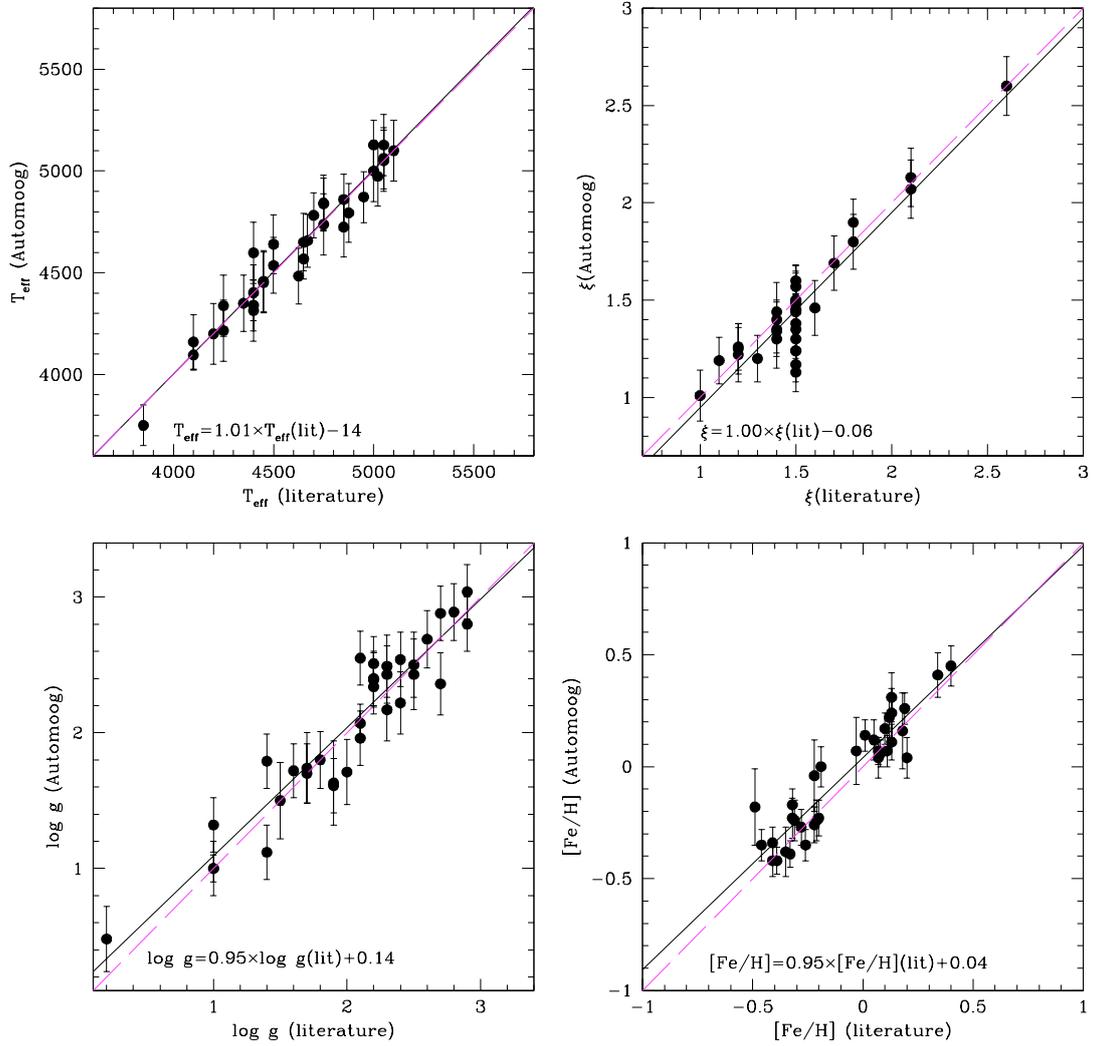}   
\caption{Comparison of the parameters derived with {\sc FAMA} with those available in the literature for the open clusters. The continuous (black) lines are  the  mean least squares fit of the {\sc FAMA} vs literature results for each of the  four parameters. The dashed (magenta) lines are the one-to-one relations for each parameter.  
 }
\label{fig_ocs}
\end{figure*}

In Table~2, we show the literature parameters of the selected stars with the errors quoted by the authors and the results obtained with our code. 
The literature parameters are given in columns 2 to 5.  
Our results are in columns 8 to 12. The iron abundances computed with  {\sc FAMA}  are re-scaled to the Solar abundances quoted in each paper (references and values in the 6th and 7th columns, respectively). 
For the  {\sc FAMA} results, the error on the iron abundance considers both  the error due to stellar parameters and the error due to the dispersion of the 
abundance around the average. 
In Fig.~\ref{fig_ocs} we plot
the literature vs. {\sc FAMA} parameters. In the four panels of  Fig.~\ref{fig_ocs} we show  the  mean least squares fit of the literature  vs. {\sc FAMA} results for each of the  four parameters with continuous (black) lines and  the one-to-one relations for each parameter with dashed (magenta) lines.  
There is  general good agreement.
Note that the correlation between the two results is close to unity for the four parameters with some small offsets.  
In particular, the offset in temperature is very small being around -$14$K, while the offset in gravity is more consistent with {\sc FAMA} tending to have higher gravity than  the literature values. This effect seems to be 
more marked at low gravity. For microturbulence the relationship is very good with a very small offset.
For [Fe/H] the results of {\sc FAMA} have an offset of 0.04~dex with respect to the literature results. 

The offsets  are  likely  due to  different aspects: for example different versions of {\sc Moog} adopted in the present work ({\sc Moog v.2010}), in 
Y12, in F10 and in M10 ({\sc Moog v.2002}), and to different model atmosphere grids. 
Y12 computed model atmospheres using the {\sc ATLAS99} program \citep{kurucz93}, F10 adopted the plane-parallel {\sc MARCS} models of \citep{bell76}, and M10 
used the Kurucz model atmospheres \citep{kurucz79}. 
On the other hand, the version of {\sc FAMA} we adopted for this test, uses the complete grid of {\sc MARCS} models \citep{gustafsson08} which includes both spherical and plane-parallel models. 
Spherical models are adopted for stars with \logg $<$3.5, thus  we practically used spherical models for the whole sample of evolved stars that we have analyzed.  
Another source of the differences with  literature results might be the result of different treatments of the line broadening. As noticed above, the results shown in the present paper have been obtained
by considering the line broadening  from collisions with neutral hydrogen for lines with tabulated damping constants \citep{barklem00, barklem05} while using the Uns\"old approximation for other lines. 
The different treatment of damping might lead to differences in abundances up to 0.1 dex.  

To probe how these  dependences work in our stars, we have analyzed one of them, namely Be18-1383,  by adopting different choices of  damping treatment, {\sc Moog} version, and  model atmospheres. 
These choices lead to the following differences: 
\begin{itemize}
\item[i)]  Uns\" old approximation for the damping treatment:  we found negligible differences in all atmospheric parameters with variation [Fe/H] on the order of $\sim$0.01~dex; 
\item[ii)] {\sc Moog} v.2009: we found differences in temperature $\sim+$20~K,  gravity $\sim$+0.04~dex,  $\xi$ $\sim$+0.05~km~s$^{-1}$, and no differences in [Fe/H];
\item[iii)] with plane-parallel Kurucz model atmospheres: we found no differences in temperature,  but differences in gravity $\sim +$0.20~dex, $\xi$ $\sim+$0.06~km~s$^{-1}$, and  [Fe/H] $\sim-$0.08~dex;
\end{itemize}

\newpage
\begin{landscape}
\begin{table*}
\caption{Sample of analyzed stars in OCs}
\begin{center}
\begin{tabular}{lllrlllllrl}
\hline\hline
Star & T$_ {\rm eff}$ & \logg\  & [Fe/H] &$\xi$ &Ref. & \fei$_{\odot}$& T$_ {\rm eff - FAMA}$ & \logg$_{\rm FAMA}$  & [Fe/H]$ _{\rm FAMA}$   &  $\xi_{\rm FAMA}$   \\
 	&    (K)       &           &          &  (km/s)          &           &           &   (K)       &            &    &  (km/s) \\  
\hline\hline
   Be31-260     &    4850$\pm$100  &         2.2$\pm$0.2 &       -0.32$\pm$0.13   &      1.5$\pm$0.2 &         F10 &         7.52 &     4860$\pm$125  &       2.34$\pm$0.20 &      -0.17$\pm$0.09   &     1.38$\pm$0.15      \\
   Be31-1295    &    5000$\pm$100  &         2.5$\pm$0.2 &       -0.32$\pm$0.16   &      1.4$\pm$0.2  &       F10   &     7.52   &     5000$\pm$150   &      2.50$\pm$0.24 &      -0.23$\pm$0.25   &      1.40$\pm$0.10     \\
      Be32-2    &    4100$\pm$100  &           1.0$\pm$0.2 &       -0.28$\pm$0.14   &      1.5$\pm$0.2  &       F10   &     7.52   &     4096$\pm$ 75   &        1.00$\pm$0.10 &      -0.27$\pm$0.08   &     1.45$\pm$0.08   \\
      Be32-4    &    4100$\pm$100  &           1.0$\pm$0.2 &       -0.31$\pm$0.14   &      1.5$\pm$0.2  &       F10   &     7.52   &     4160$\pm$134   &     1.32$\pm$0.20 &      -0.24$\pm$0.09   &     1.48$\pm$0.12   \\
      Be39-3    &    4200$\pm$100  &         1.5$\pm$0.2 &        -0.20$\pm$0.12   &      1.5$\pm$0.2  &       F10   &     7.52   &     4200$\pm$150   &      1.50$\pm$0.28 &      -0.23$\pm$0.09   &      1.50$\pm$0.15      \\
      Be39-5    &    4450$\pm$100  &         1.8$\pm$0.2 &       -0.21$\pm$0.15   &      1.5$\pm$0.2  &       F10   &     7.52   &     4452$\pm$150   &      1.80$\pm$0.21 &      -0.24$\pm$0.09   &     1.57$\pm$0.11     \\
     Be39-12    &    4750$\pm$100  &         2.2$\pm$0.2 &       -0.19$\pm$0.18   &      1.5$\pm$0.2  &       F10   &     7.52   &     4838$\pm$128   &     2.51$\pm$0.20 &       0.00$\pm$0.11   &     1.24$\pm$0.10    \\
     Be39-14    &    4750$\pm$100  &         2.2$\pm$0.2 &       -0.22$\pm$0.16   &      1.5$\pm$0.2  &       F10   &     7.52   &     4842$\pm$138   &      2.40$\pm$0.20 &      -0.04$\pm$0.17   &      1.30$\pm$0.10     \\
     M67-105    &    4400$\pm$100  &           2.0$\pm$0.2 &        0.01$\pm$0.13   &      1.5$\pm$0.2  &       F10   &     7.52   &     4314$\pm$150   &     1.71$\pm$0.24 &       0.14$\pm$0.08   &     1.17$\pm$0.09      \\
     M67-141    &    4700$\pm$100  &         2.4$\pm$0.2 &         0.10$\pm$0.13   &      1.5$\pm$0.2  &       F10   &     7.52   &     4782$\pm$110   &     2.54$\pm$0.20 &       0.17$\pm$0.08   &     1.49$\pm$0.13      \\
     M67-170    &    4250$\pm$100  &         1.4$\pm$0.2 &       -0.03$\pm$0.14   &      1.5$\pm$0.2  &       F10   &     7.52   &     4338$\pm$150   &     1.79$\pm$0.20 &       0.07$\pm$0.16   &     1.44$\pm$0.11     \\
  NGC188-532    &    4850$\pm$100  &         2.9$\pm$0.2 &        0.18$\pm$0.2   &      1.5$\pm$0.2  &       F10   &     7.52   &     4724$\pm$146   &      2.80$\pm$0.20 &       0.16$\pm$0.17   &     1.24$\pm$0.09     \\
  NGC188-747    &    4650$\pm$100  &         2.6$\pm$0.2 &        0.13$\pm$0.2   &      1.5$\pm$0.2  &       F10   &     7.52   &     4448$\pm$141   &     2.19$\pm$0.21 &       0.32$\pm$0.12   &     0.96$\pm$0.10    \\
  NGC188-919    &    4400$\pm$100  &         2.1$\pm$0.2 &        0.13$\pm$0.2   &      1.5$\pm$0.2  &       F10   &     7.52   &     4598$\pm$150   &      2.55$\pm$0.20 &       0.24$\pm$0.11   &     1.50$\pm$0.14     \\
 NGC188-1224    &    4750$\pm$100  &         2.8$\pm$0.2 &        0.12$\pm$0.16   &      1.5$\pm$0.2  &       F10   &     7.52   &     4738$\pm$150   &     2.89$\pm$0.21 &       0.22$\pm$0.09   &     1.35$\pm$0.13   \\
 NGC1193-282    &    4650$\pm$100  &         2.1$\pm$0.2 &       -0.22$\pm$0.14   &      1.5$\pm$0.2  &       F10   &     7.52   &     4568$\pm$ 98   &     1.96$\pm$0.20 &      -0.26$\pm$0.08   &     1.45$\pm$0.13      \\
   Be18-1163    &    4500$\pm$100  &         2.2$\pm$0.3 &       -0.46$\pm$0.10   &      1.2$\pm$0.2  &       Y12   &      7.5   &     4640$\pm$144   &     2.39$\pm$0.20 &      -0.35$\pm$0.07   &     1.22$\pm$0.14     \\
   Be18-1383    &    4400$\pm$100  &         1.9$\pm$0.3 &       -0.41$\pm$0.12   &      1.2$\pm$0.2  &       Y12   &      7.5   &     4402$\pm$138   &     1.63$\pm$0.31 &      -0.42$\pm$0.09   &     1.26$\pm$0.12      \\
    Be21-T50    &    4625$\pm$100  &         1.9$\pm$0.3 &       -0.26$\pm$0.13   &      1.3$\pm$0.2  &       Y12   &      7.5   &     4484$\pm$137   &     1.61$\pm$0.20 &      -0.35$\pm$0.08   &      1.20$\pm$0.12      \\
    Be21-T51    &    4500$\pm$100  &         1.7$\pm$0.3 &       -0.35$\pm$0.14   &      1.2$\pm$0.2  &       Y12   &      7.5   &     4536$\pm$137   &     1.74$\pm$0.26 &      -0.38$\pm$0.12   &     1.25$\pm$0.13     \\
    Be22-414    &    4350$\pm$100  &         1.7$\pm$0.3 &       -0.41$\pm$0.11   &      1.1$\pm$0.2  &       Y12   &      7.5   &     4350$\pm$139   &      1.70$\pm$0.22 &      -0.34$\pm$0.07   &     1.19$\pm$0.12     \\
    Be22-643    &    3850$\pm$100  &         0.2$\pm$0.3 &       -0.49$\pm$0.22   &      1.5$\pm$0.2  &       Y12   &      7.5   &     3750$\pm$100   &     0.48$\pm$0.24 &      -0.18$\pm$0.23   &      1.60$\pm$0.08     \\
     Be32-16    &    4875$\pm$100  &         2.4$\pm$0.3 &       -0.39$\pm$0.11   &        1.0$\pm$0.2  &       Y12   &      7.5   &     4794$\pm$144   &     2.22$\pm$0.23 &      -0.42$\pm$0.10   &     1.01$\pm$0.13   \\
     Be32-18    &    4950$\pm$100  &         2.7$\pm$0.3 &       -0.33$\pm$0.10   &      1.4$\pm$0.2  &       Y12   &      7.5   &     4872$\pm$125   &     2.36$\pm$0.23 &      -0.39$\pm$0.08   &     1.35$\pm$0.14      \\
   NGC6192-9    &    5050$\pm$ 70  &         2.3$\pm$.15 &        0.19$\pm$0.07   &      1.7$\pm$0.25  &       M10   &     7.47   &     5128$\pm$150   &     2.49$\pm$0.23 &       0.26$\pm$0.10   &     1.69$\pm$0.14      \\
  NGC6192-45    &    5020$\pm$ 70  &         2.5$\pm$.15 &        0.08$\pm$0.08   &      1.6$\pm$0.25  &       M10   &     7.47   &     4974$\pm$147   &     2.43$\pm$0.26 &       0.06$\pm$0.10   &     1.46$\pm$0.14   \\
  NGC6192-96    &    5050$\pm$ 70  &         2.3$\pm$.15 &        0.13$\pm$0.10   &      2.1$\pm$0.25  &       M10   &     7.47   &     5050$\pm$150   &     2.43$\pm$0.21 &       0.11$\pm$0.12   &     2.13$\pm$0.15     \\
 NGC6192-137    &    4670$\pm$ 70  &         2.1$\pm$.15 &        0.07$\pm$0.08   &      1.8$\pm$0.25  &       M10   &     7.47   &     4658$\pm$130   &     2.07$\pm$0.14 &       0.04$\pm$0.09   &      1.80$\pm$0.14     \\
   NGC6404-5    &    5000$\pm$ 70  &           1.0$\pm$.15 &        0.05$\pm$0.09   &      2.6$\pm$0.25  &       M10   &     7.47   &     5128$\pm$120   &        1.00$\pm$0.20 &       0.12$\pm$0.09   &      2.60$\pm$0.15      \\
  NGC6404-16    &    4450$\pm$ 70  &         1.6$\pm$.15 &        0.07$\pm$0.09   &      2.1$\pm$0.25  &       M10   &     7.47   &     4458$\pm$150   &     1.72$\pm$0.20 &       0.07$\pm$0.06   &     2.07$\pm$0.15   \\
  NGC6404-27    &    4400$\pm$ 70  &         1.4$\pm$.15 &         0.20$\pm$0.09   &      1.8$\pm$0.25  &       M10   &     7.47   &     4340$\pm$125   &     1.12$\pm$0.20 &       0.04$\pm$0.09   &      1.90$\pm$0.12     \\
  NGC6404-40    &    4250$\pm$ 70  &         2.3$\pm$.15 &        0.11$\pm$0.10   &      1.4$\pm$0.25  &       M10   &     7.47   &     4216$\pm$150   &     2.17$\pm$0.23 &       0.07$\pm$0.08   &     1.34$\pm$0.11      \\
  NGC6583-46    &    5100$\pm$ 70  &         2.9$\pm$.15 &         0.4$\pm$0.12   &      1.4$\pm$0.25  &       M10   &     7.47   &     5100$\pm$150   &     3.04$\pm$0.20 &       0.45$\pm$0.09   &     1.44$\pm$0.15   \\
  NGC6583-62    &    5050$\pm$ 70  &         2.7$\pm$.15 &        0.34$\pm$0.12   &      1.4$\pm$0.25  &       M10   &     7.47   &     5062$\pm$150   &     2.88$\pm$0.20 &       0.41$\pm$0.10   &      1.30$\pm$0.15   \\
      \hline
\hline
\end{tabular}
\label{tab_oc}
\end{center}
\end{table*}
\end{landscape}

\subsection{Stars in Globular Clusters} 
We have selected a sample of stars that belong to GCs observed by \citet{carretta09} (C09, hereafter)  and span a wide metallicity range ([Fe/H] from  -0.8 to -2.3) to check the applicability of our method  in the metal poor 
regime.  The selected stars\footnote{The EW measurements have been kindly provided by E. Carretta togheter with  the adopted atomic data} belong to the clusters 47Tuc, M4, M10, M15, and NGC6397. The literature values of  temperature and gravity  of these stars are derived from photometry, and thus,  a direct comparison with the parameters obtained from our spectral analysis is not possible, with the exception of metallicity.  The only free parameter in C09 is 
the microturbolent velocity and, of course, the metallicity.  
We mention that the group of E. Carretta uses a private line analysis code, known as the {\sc ROSA} code \citep{gratton88}, which produces solid results in accord with our work.

As in our test on OC stars, we have started the analysis with {\sc FAMA} from 
the photometric values of T$_{\rm eff}$ and \logg\ and from the values of $\xi$ given in C09. 
In Table~\ref{table_gc}, we show these values of  T$_{\rm eff}$, \logg,  [Fe/H], and  $\xi$. 
Then  we report the results of our spectroscopic analysis in the following columns, with  T$_{\rm eff}$, \logg, [Fe/H], and $\xi$ derived using {\sc FAMA}, which is 
rescaled to the Solar abundance adopted by C09. 

\begin{figure}
\centering
\includegraphics[width=0.5\textwidth]{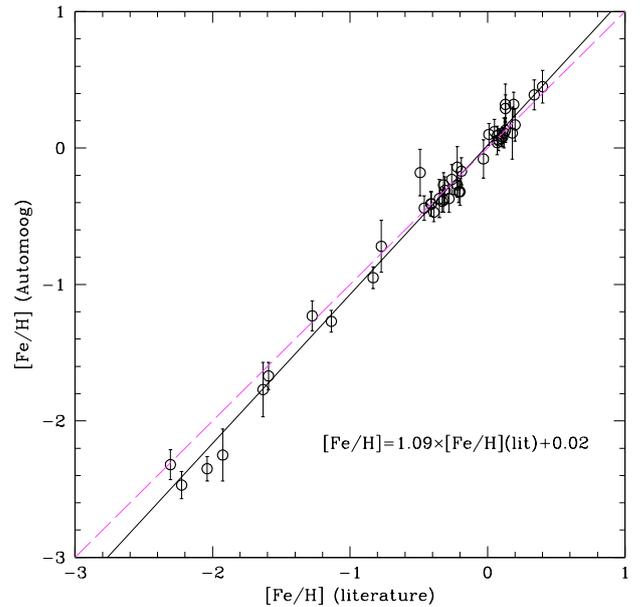}   
\caption{Comparison of the metallicity derived with {\sc FAMA} to the literature values for the sample of OC and GC stars.  The continuous (black) lines are  the  mean least squares fit of the {\sc FAMA} vs literature results for each of the  four parameters. The dashed (magenta) lines are the one-to-one relations for each parameter.  
 }
\label{fig_cs}
\end{figure}

Figure~\ref{fig_cs} summarizes the literature vs. {\sc FAMA} metallicity for both OCs and GCs, 
and shows the good agreement over the whole metallicity range: The coefficient of the mean least squares fit is very close to unity, at 1.09.

\begin{table*}
\caption{Sample of analyzed stars in GCs}
\begin{center}
\begin{tabular}{lllrlllrl}
\hline\hline
Star & T$_ {\rm eff}$ & \logg\  & [Fe/H] &$\xi$ &T$_ {\rm eff - FAMA}$ & \logg$_{\rm FAMA}$  & [Fe/H]$ _{\rm FAMA}$   &  $\xi_{\rm FAMA}$   \\
 	&    (K)       &           &          &  (km/s)          &         (K)       &            &    &  (km/s) \\  
\hline\hline
47Tuc-5270 &    3999 &   1.01 &   -0.772 & 1.48 & 3909$\pm$175 & 1.01$\pm$0.40 & -0.72$\pm$0.19 & 1.48$\pm$0.1 \\
47Tuc-13795 &    4183 &   1.30 &   -0.832 & 1.49 & 4183$\pm$150 & 1.11$\pm$0.20 & -0.95$\pm$0.08 & 1.74$\pm$0.1 \\
M10-28853 &    4381 &   1.29 &   -1.593 & 1.66 & 4407$\pm$150 & 1.24$\pm$0.20 & -1.67$\pm$0.10 & 1.97$\pm$0.1 \\ 
M10-28854 &    4425 &   1.36 &   -1.633 & 1.70 & 4447$\pm$75  & 1.10$\pm$0.20 & -1.77$\pm$0.20 & 1.62$\pm$0.15 \\
M4-27448 &    4243 &   1.24 &   -1.136 & 1.97 & 4207$\pm$75  & 1.00$\pm$0.10 & -1.27$\pm$0.08 & 1.66$\pm$0.07 \\ 
M4-29693 &    4261 &   1.28 &   -1.275 & 1.66 & 4489$\pm$150 & 1.83$\pm$0.20 & -1.23$\pm$0.11 & 1.98$\pm$0.1 \\  
NGC6397-602241 &    4779 &   1.69 &   -1.926 & 1.50 & 4711$\pm$125 & 1.14$\pm$0.20 & -2.25$\pm$0.19 & 1.99$\pm$0.1 \\ 
NGC6397-602256 &    4720 &   1.57 &   -2.039 & 2.16 & 4660$\pm$75  & 1.00$\pm$0.10 & -2.35$\pm$0.09 & 2.16$\pm$0.07 \\
M15-2792 &    4567 &   1.26 &   -2.306 & 1.88 & 4729$\pm$150 & 1.35$\pm$0.20 & -2.32$\pm$0.11 & 1.50$\pm$0.1 \\ 
M15-4099 &    4324 &   0.69 &   -2.225 & 2.16 & 4324$\pm$75  & 0.68$\pm$0.10 & -2.47$\pm$0.10 & 2.16$\pm$0.07 \\
      \hline
\hline
\end{tabular}
\label{table_gc}
\end{center}
\end{table*}

\section{Summary and conclusions}

We have presented a new  {\sc FAMA} code designed for an automatic spectral analysis using EWs obtained from  high-resolution stellar spectra.
The code is written in Perl and is based in the widely used {\sc fortran} code,  {\sc Moog}, by C. Sneden. {\sc FAMA} is freely distributed (via email to {\tt laura@arcetri.astro.it}, web page {\tt http//www.arcetri.astro.it/$\sim$laura}, and CDS) and it works on different platforms (tested on {\sc Mac osx} (Leopard, Mountain Lion) and {\sc Linux} (Ubuntu, Centos)). 

In the present paper, we have described the code with its general structure, starting from the 
determination of the atmospheric parameters that are defined by the excitation and ionization equilibria, and by the nullification 
of the trend between \fei\ and reduced EWs. We then described how metallicity is  computed, and how errors 
on stellar parameters and on abundances are obtained.
We have also described how the convergence criteria are set and how they are chosen and varied depending on the quality of the spectra and of the EW measurements. 

We have presented three tests: the first one on the reproducibility of the final results, analyzing the Solar spectrum starting from 
six different starting points,  the second one computing a complete spectral analysis of a set of    34 evolved stars belonging to Galactic OCs, and the third one 
computing the spectral analysis of 10 stars belonging to GCs with literature metallicity [Fe/H] values that range from -0.7 to -2.3. 
To separate the effects due to EW measurements from the choice of atomic parameters, we have used published EWs and log {\em gf}  for all tests. 
We have demonstrated the independence of {\sc FAMA} on the initial set of stellar parameters, and its capability to reproduce the 
literature results obtained with a completely manual and time-consuming technique. 
Small offsets and differences with literature OC results might be due to different versions of {\sc Moog},  to a different adopted  grid of model atmospheres,
and to line broadening treatment.  In GCs, we could compare [Fe/H] to literature values, finding a good agreement, even if we have  a slightly higher dispersion than in more metal-rich stars.  

{\sc FAMA} is an easily installable  Perl code, which performs a complete spectral analysis of high-resolution spectra. It starts from two files with EWs
 and a file with first-guess atmospheric parameters and gives a file with stellar parameters with their uncertainties,  a file with the element abundances with their errors, and  
 a control plot to check the final solution as main final results. 
 Due to its versatility and its low CPU requirement\footnote{For a typical stars with $\sim$150-200 \fei\ lines and $\sim$20-30 \feii\ lines the CPU time to complete the analysis  on a iMac  (Processor 2.9 GHz, RAM 16 GB) is about 2 minutes}, it is a perfect code to perform spectral analysis on large samples of spectra.

\begin{acknowledgements}
We warmly thank the referee, Chris Sneden, for his support to the paper and for his useful comments and suggestions which helped us to improve the paper. 
We thank Eugenio Carretta for providing us the EWs of the sample of GC stars. We finally thank the language editor, C.-J. Lin, for improving the quality of our English. This research has made use of NASA's Astrophysics Data System.
 \end{acknowledgements}

\end{document}